\documentclass[runningheads,a4paper]{llncs}

\usepackage{amssymb}
\usepackage{graphicx}

\usepackage{booktabs} % For formal tables
\usepackage{graphicx}
\usepackage{subfigure}
\usepackage{units}
\usepackage{blindtext}
\usepackage{cite}
\usepackage{xcolor}
\usepackage{amsmath}
\usepackage{paralist}

\usepackage{fancyhdr}

\usepackage[ruled]{algorithm2e}
\usepackage{algorithmic}
\LinesNumbered

\usepackage{url}

\usepackage[
    n,
    advantage,
    operators,
    sets,
    adversary,
    landau,
    probability,
    notions,
    logic,
    ff,
    mm,
    primitives,
    events,
    complexity,
    asymptotics,
    keys]{cryptocode}

\newcommand\KRBCCN{\texttt{KRB-CCN}}
\newcommand\KTGS{\texttt{CGT-Prod}}
\newcommand\KAS{\texttt{TGT-Prod}}

\newtheorem{cl}{\textbf{Claim}}
\newtheorem{prop}{\textbf{Property}}
\newtheorem{ps}{\textbf{Proof (sketch)}}

\usepackage{url}
\begin{document}

\mainmatter  % start of an individual contribution

% first the title is needed
\title{\KRBCCN: Lightweight Authentication \& Access Control for Private Content-Centric Networks}

% a short form should be given in case it is too long for the running head
\titlerunning{\KRBCCN: Lightweight Authentication \& Access Control for Private CCNs}

% the name(s) of the author(s) follow(s) next
%
% NB: Chinese authors should write their first names(s) in front of
% their surnames. This ensures that the names appear correctly in
% the running heads and the author index.
%
\author{Ivan O. Nunes \and Gene Tsudik}
\authorrunning{Nunes and Tsudik}
% (feature abused for this document to repeat the title also on left hand pages)

% the affiliations are given next; don't give your e-mail address
% unless you accept that it will be published
\institute{University of California Irvine, USA\\
\{ivanoliv,g.tsudik\}@uci.edu}

%
% NB: a more complex sample for affiliations and the mapping to the
% corresponding authors can be found in the file "llncs.dem"
% (search for the string "\mainmatter" where a contribution starts).
% "llncs.dem" accompanies the document class "llncs.cls".
%

%\toctitle{Lecture Notes in Computer Science}
%\tocauthor{Authors' Instructions}
\maketitle

\begin{abstract}
Content-Centric Networking (CCN) is an internetworking paradigm that offers an alternative 
to today's IP-based Internet Architecture.  Instead of focusing on hosts and their locations, CCN 
emphasizes addressable named content. By decoupling content from its location, CCN allows 
opportunistic in-network content caching, thus enabling better network utilization, at least for
scalable content distribution. 
However, in order to be considered seriously, CCN must support basic security services, including 
content authenticity, integrity, confidentiality, authorization and access control. Current approaches rely on
content producers to perform authorization and access control, which is typically attained via  
public key encryption. This general approach has several disadvantages.  First, consumer privacy vis-a-vis 
producers is not preserved. Second,  identity management and access control impose high computational overhead
on producers. Also, unnecessary repeated authentication and access control decisions must be made for each 
content request. (This burden is particularly relevant for resource-limited producers, e.g., anemic IoT devices.)

These issues motivate our design of \KRBCCN\ -- a complete authorization and access control system for private
CCN networks. Inspired by Kerberos in IP-based networks, \KRBCCN\ 
involves distinct authentication and authorization authorities. By doing so, \KRBCCN\ obviates the need for 
producers to make consumer authentication and access control decisions. \KRBCCN\ preserves consumer privacy
since producers are unaware of consumer identities. Producers are also not required to keep any hard state 
and only need to perform two symmetric key operations to guarantee that sensitive content is confidentially 
delivered only to authenticated and authorized consumers. Furthermore, \KRBCCN\ works transparently on the consumer 
side. Most importantly, unlike prior designs, \KRBCCN\ leaves the network (i.e., CCN routers) out of any
authorization, access control or confidentiality issues.  
We describe \KRBCCN\ design and implementation, analyze its security, and report on its performance.\\

\begin{footnotesize}
\textit{\textbf{Note:} This is the extended version of a manuscript published with the same title at ACNS 2018 - Applied Cryptography \& Network Security.}
\end{footnotesize}
\end{abstract}

\section{Introduction}
Content-Centric Networking (CCN) is an emerging internetworking paradigm that emphasizes transfer
of named data (aka content) instead of host-to-host communication~\cite{jacobson2009networking,zhang2010named}. 
All CCN content is uniquely named. Content \emph{producers} are entities that publish content under namespaces.
Entities that wish to obtain content, called \emph{consumers}, do so by issuing an \emph{interest} message 
specifying desired content by its unique name. The network is responsible for
forwarding the interest, based on the content name, to the nearest copy of requested content. 
Interests do not carry source or destination addresses. Each interest leaves state in every router it traverses.
This state is later used to forward, along the reverse path, requested \emph{content} back to the consumers.
As content is forwarded to the consumer, each router can choose to cache it. If a popular content is cached, 
subsequent interests for it can be satisfied by the caching router and not forwarded further. 
This can lead to lower delays, better throughput and improved network utilization.

Due to CCN's unique characteristics, security focus shifts from 
securing host-to-host tunnels to securing the content itself. CCN mandates that each 
content must be signed by its producer. This is the extent of CCN network-layer security.
In particular, CCN does not make any provisions for confidentiality, authorization or access
control, leaving these issues to individual applications.
We believe that this approach makes sense, since involving the network (i.e., routers) in such issues
is generally problematic for both performance and security reasons.

Access control (AC) in CCN has been explored in recent years. Most approaches
~\cite{smetters2010ccnx, misra2013secure, wood2014flexible, ion2013toward, kuriharay2015encryption, 
yu2015name, ghali2015interest}
rely on using public key encryption to safeguard content (we overview these approaches in Section~\ref{rw}). 
Specifically, producers are expected to encrypt content 
with a public key corresponding to an authorized consumer, or a group thereof. 
The latter use their corresponding private keys to decrypt. Although it seems to work, 
this approach exhibits several problems:
\begin{compactitem}
\item First, producers are responsible for handling consumer authentication and content AC on their own.
Thus, they must deal with (1) consumer identity management and authentication, (2) AC policy representation 
and storage, (3) updates of access rights, and (4) content encryption. In some cases, producers might not want
(or be able) to deal with this burden, e.g.,  resource-constrained IoT devices. On the consumer side, this means
keeping track of producer-specific authentication contexts and keys.
\item Second, AC enforced by producers implies sacrificing consumer privacy, which is an important
and appealing CCN feature. Since CCN interests do not carry source addresses, a content producer 
(or a router) normally does not learn the identity of the consumer. However, if the producer enforces AC,
it learns consumers identities.
\item Third, if a set of producers belong to the same administrative domain
and each producer enforces its AC policy, it is difficult to react to policy changes,  
e.g., access revocation for a given consumer or a consumer's credential.
Implementing such changes requires notifying each producer individually.
\item Finally, since public keys bind authorization rules to identities, authentication of consumers 
is attained via consumer-owned private keys. However, if consumer is authenticated by other means,
e.g., passwords and biometrics,  each producer would have to store and manage potentially sensitive
state information (password files or biometric templates) for each consumer.
\end{compactitem}
Since mid-1980s, Kerberos~\cite{neuman1994kerberos} has been successfully and widely used  
to address these exact issues in private IP-based networks or so-called stub Autonomous Systems.
Kerberos de-couples authentication and authorization services
via short-term {\em tickets}. It also allows services (e.g., storage, compute or web servers) to
be accessed by clients over a secure ephemeral session. By checking a client's ticket for freshness
of authentication information, a service limits the period of vulnerability due to revocation.

In this paper, we present \KRBCCN, a system inspired by Kerberos for authentication and access control (AC)
enforcement in CCN, that aims at addressing the aforementioned issues \footnote{\KRBCCN\ source-code is available at: \url{https://github.com/ivanolive/krb-ccn}}. 
\KRBCCN\ treats consumer authentication and authorization as
separate services. It uses tickets to allow consumers to convey authorization
permissions to  servers, e.g., content producers or repositories. Servers use tickets
to determine whether requested content should be provided. \KRBCCN\ also introduces a
novel namespace based AC policy, which allows a consumer to securely retrieve content without 
revealing its identity to the content producer, thus preserving consumer privacy.
In addition, \KRBCCN\ is transparent to the users; they need not be aware of \KRBCCN\ or perform any 
additional tasks. It is also completely invisible to the network, i.e., CCN routers are unaware of \KRBCCN.

\textbf{Organization:} Section~\ref{background} overviews CCN and Section~\ref{kerberos} overviews Kerberos.
Next, Section~\ref{design} introduces \KRBCCN, including its system architecture, namespace based AC scheme, 
and the protocol for authentication, authorization, and secure content retrieval.
In addition, a security analysis for the design presented in Section~\ref{design} is provided in Appendix A.
%Section~\ref{security} discusses \KRBCCN\ security as well as its advantages and limitations compared to other approaches.
Then, performance of \KRBCCN\ is evaluated in Section~\ref{eval}. Finally, Section~\ref{rw} discusses related work and 
Section~\ref{sec:conclusion} concludes this paper.

\section{CCN Overview}\label{background}
We now overview key features of CCN. Given basic familiarity with CCN, this section
can be skipped with no loss of continuity.

In contrast to today's IP-based Internet architecture which focuses on end-points of communication
(i.e., interfaces/hosts and their addresses) CCN \cite{jacobson2009networking,mosko2016semantics} 
centers on content by making it named, addressable, and routable within the network. Moreover, a content
must be signed by its producer. A content name is 
a URI-like string composed of one or more variable-length name segments, separated by the \url{`/'} character. 
To obtain content, a user (consumer) issues an explicit request message, called an \emph{interest}
containing the name of desired content. This interest can be \emph{satisfied} by either: (1) a router cache, or (
2) the content producer. A \emph{content object} message is returned to the consumer upon satisfaction of
the interest. Name matching is exact, e.g., an interest for \url{/edu/uni-X/ics/cs/fileA} can only be satisfied by 
a content object named \url{/edu/uni-X/ics/cs/fileA}.

In addition to a payload, a content object includes several other fields. In this paper,
we are only interested in the following three: {\tt Name}, {\tt Validation}, and {\tt ExpiryTime}.
{\tt Validation} is a composite of validation algorithm information
(e.g., the signature algorithm, its parameters, and the name of the public
verification key), and validation payload, i.e., the content signature. We use the
term ``signature'' to refer to the entire {\tt Validation} field. {\tt ExpiryTime} is an optional,
producer-recommended duration for caching a content object.
Interest messages carry a name, optional payload, and other fields that restrict the 
content object response. We refer to \cite{mosko2016semantics} for a complete description 
of all CCN message types, fields and their semantics.

Packets are moved within the network by routers. Each CCN router has two mandatory (and one
optional) components:
\begin{compactitem}
\item {\em Forwarding Interest Base} (FIB) -- a table of name prefixes and
  corresponding outgoing interfaces. The FIB is used to route interests based on
  longest-prefix-matching (LPM) of their names.
\item {\em Pending Interest Table} (PIT) -- a table of outstanding (pending)
  interests and a set of corresponding incoming interfaces. 
\item An optional {\em Content Store} (CS) used for content
 caching. The timeout for cached content is specified in the \texttt{ExpiryTime}
 field of the content header. From here on, we use the terms {\em CS} and
{\em cache} interchangeably.
\end{compactitem}
A router uses its FIB to forward interests toward the producer of requested content.
Whereas, a router uses its PIT to forward content along the reverse path towards consumers. 
Specifically, upon receiving an interest, a router $R$ first checks its cache
(if present) to see if it can satisfy this interest locally. In case of a cache miss, 
$R$ checks its PIT for an outstanding version of the same interest. If there is a PIT match, 
the new interest's incoming interface is added to the PIT entry. Otherwise, $R$ creates a new PIT
entry and forwards the interest to the next hop according to its FIB (if possible). For each forwarded interest, 
$R$ stores some state information in the PIT entry, including the name in the interest and the
interface from which it arrived, such that content may be returned to the
consumer. When content is returned, $R$ forwards it to all interfaces listed in
the matching PIT entry and then removes the entry. A content 
that does not match any PIT entry is discarded. 

\section{Kerberos Overview}\label{kerberos}
We now summarize Kerberos. We refer to~\cite{krb_protocol_tutorial} for a more extensive description.
Kerberos includes four types of entities: clients, services, an Authentication Server (AS), and a 
Ticket-Granting Server (TGS). The AS/TGS pair (which are often collocated within the same host) 
is also known as a Key Distribution Center (KDC). Should a new client/user or a new service be added 
to the network, it must first be properly registered into KDC's (AS and TGS) databases.

%%
%%Ivan: I am removing Autonomous System (AS) abreviation, because AS is already used for "Authentication Server".
%%
In Kerberos' terminology, a \textbf{\textit{realm}} corresponds to a single administrative domain, 
e.g., a private network or a stub Autonomous System. Each realm has one KDC and 
any authorization or AC decision by a KDC is only valid within its realm.
Thus, identities, tickets, and encryption keys (see below) are also realm-specific.

\textbf{\textit{Principal}} is the term used to refer to names of entries in the KDC database. 
Each user/client and service has an associated principal. User principals are generally their 
usernames in the system. Service principals are used to specify various applications.
A service principal has the format: \url{service/hostname@realm}. A service specification 
is needed in addition to a hostname, since a single host often runs multiple services.
With Kerberos operation in IP Networks, principals are resolved to host IP addresses via DNS 
look-ups~\cite{mockapetris1987domain}. As can be expected, CCN obviates the need for 
%%GTS: not sure about that... How do your discover services?
%%
%%Ivan (Ans): Search engine or websites' links (www or intranet websites). 
% But name to host-addr DNS queries aren't necessary.
%%	      We can imagine that content name are equivalent to current URIs.
DNS look-ups, since all content objects are uniquely named by design.
Moreover, routing is done based on content names. As discussed below, 
\KRBCCN\ enforces AC based on content namespaces, instead of service principals.

Each client/user principal (i.e., username) stored in the AS database is associated with a key,
which can be either a public-key or a symmetric key derived from the user's password.
Also, the same client/user principal also exists in the TGS database. However, it is associated 
with a list principals for services that such user has permission to access.

Before attempting to access any content, a client must first authenticate to its local AS.
This is done by either typing a password, or proving possession of a secret key associated with 
the client's identity in the AS database. If the client proves its identity, AS issues a 
\textbf{\textit{Ticket-Granting Ticket (TGT)}} -- a temporary proof of identity required for the authorization.
This TGT might be cached and used multiple times until its expires.

The client uses a valid TGT to request, from TGS, authorization for a service.
The TGS is responsible for access control decisions -- verifying whether the requested service is within 
the set of permitted services for the identity ascertained in the provided TGT. If the result is positive, 
TGS issues a \textbf{\textit{Service Ticket (ST)}} to be used for requesting the actual service.

%%GTS: This is a poorly worded paragraph. Does it say anything useful? Delete?
%Proper encryption of tickets (TGT and ST) and effective distribution of session keys (session keys are 
%provided to clients and services together with the tickets to enable secure communication of data between them) 
%guarantee the security of authentication, authorization, and data confidentiality.

\section{\KRBCCN\ Design}\label{design}
There are three fundamental requirements for any authentication and authorization system. First, 
AC policies must effectively bind identities to their access rights. Second, once AC policies are established, 
there must be a way to enforce them, thus preventing unauthorized access. Third, authentication 
mechanisms must ensure that identities can not be spoofed; this includes both producers and consumers.
The system must also not involve the network elements (i.e., routers) where authentication and authorization 
burden is both misplaced and simply unnecessary.

In the rest of this section, we describe how \KRBCCN\ achieves each of these requirements. We start by 
introducing \KRBCCN\ system architecture and its namespace-based AC policy, which takes advantage of 
CCN hierarchical content name structure to provide AC based on content prefixes. Next, we describe 
\KRBCCN\ communication protocol, which enforces AC policies while providing a single sign-on mechanism 
for user authentication. Though \KRBCCN\ is inspired by Kerberos for IP-based networks, 
it also takes advantage of unique CCN features to effectively satisfy basic 
authentication and authorization requirements. Throughout the protocol description we discuss the intuition
behind the security of \KRBCCN. A detailed security analysis of \KRBCCN\ is provided in Appendix A.

As it is the case for IP-based Kerberos, \KRBCCN\ targets private (content-centric) networks,
such as intra-corporation/intra-Autonomous Systems settings.

\begin{figure}[!h]
\begin{center}
\includegraphics[width=0.5\columnwidth]{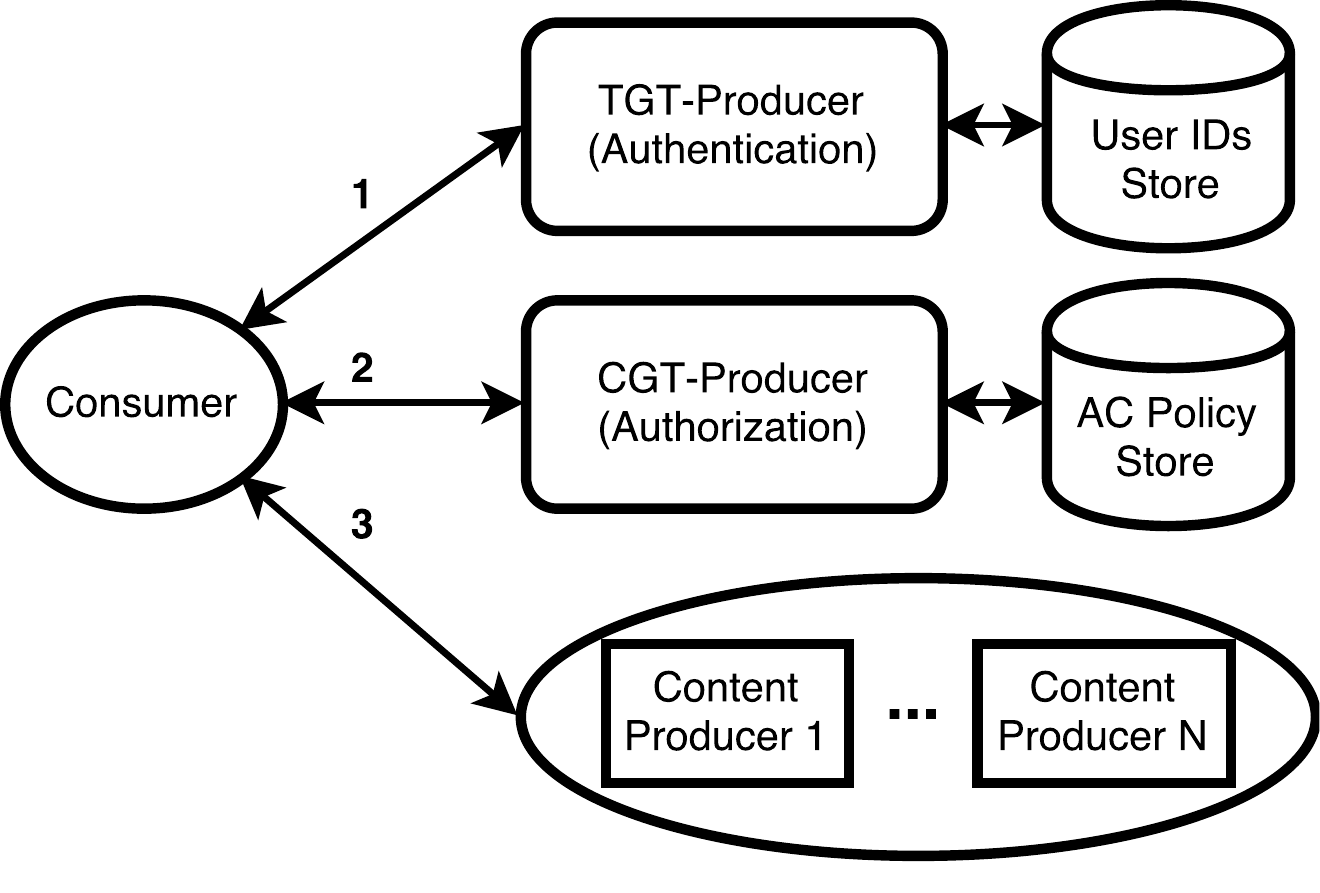}
\caption{\KRBCCN\ system architecture}
\label{arch}
\end{center}
\end{figure}

\subsection{System Architecture}
Recall that Kerberos has 4 types of entities: AS, TGS, client, and server.
A Kerberos realm (domain) typically has one AS/TGS pair, usually collocated in the same 
host, as well as multiple clients and servers.
\KRBCCN\ also includes four types of entities that map to Kerberos entities as follows: 
\begin{itemize}
\item Consumer: corresponds to Kerberos client. It issues interests 
for content and services according to \KRBCCN\ protocol. Each consumer has an identity and a set of 
associated permissions registered in the system. 
\item Producer: subsumes one or more Kerberos servers. A producer is required to register 
its namespace(s) in the system, by registering with a TGS (see below). A single namespace, i.e., a name prefix,
can correspond to a single Kerberos server. Alternatively, a group of namespaces of the same producer
can be treated as a single server. A producer does not perform any direct consumer authentication or authorization.
A producer only checks whether a content-requesting consumer possesses a valid TGS-issued ticket.
\item As in Kerberos, authentication and authorization are handled by two logically separated entities
which can be collocated: 

\noindent {\em\underline{Authentication Server (AS) (aka \KAS): }} treated as a special type of producer that 
generates so-called Ticket-Granting Tickets (TGT-s) for consumers once their identities are
verified. These tickets can then be used as temporary proofs of identity. We refer to the 
AS as \emph{\KAS}.

\noindent {\em\underline{Ticket-Granting Server (TGS) (aka \KTGS): }} performs authorization and is 
also treated as a special type of producer. Based on a valid consumer TGT and a request to 
access to a given namespace (server), TGS checks whether this consumer
is allowed access to the requested namespace. If so, TGS issues a Content-Granting Ticket (CGT), which 
proves to the producer that this consumer is granted access to any content 
under the requested namespace. We refer to TGS as \emph{\KTGS}.
\end{itemize}
Figure~\ref{arch} illustrates \KRBCCN\ system architecture. As part of the log-in procedure (aka single sign-on or SSO),
a consumer authenticates to \KAS\ (round 1) and obtains a TGT, which it caches. Whenever a 
consumer wants to initially access content from a particular producer, it requests authorization from 
\KTGS\ using its cached TGT (round 2) and obtains a CGT, which it also caches. 
A CGT authorizes access to one or more namespaces belonging to the same producer.
Finally, a consumer requests content from the producer using the corresponding CGT (round 3).
TGT and CGT-s remain valid and re-usable until their expiration time runs out.
Note that each round (1, 2 and 3) is realized as a single interest-content exchange.

Subsequent requests from the namespace(s) specified in the CGT require no involvement of either 
\emph{\KAS} or \emph{\KTGS}. A consumer retrieves another content  
by directly issuing an interest containing the cached CGT. To access content under a different namespace,
a consumer uses its cached TGT to contact \KTGS\ and request a new CGT. 
% Therefore, requests under different namespaces can be done in two communication rounds, 
% using TGT as an authentication cookie for single-sign-on.

For authentication and authorization, \KRBCCN\ must ensure that TGT-s and 
CGT-s issued to a specific consumer $C_r$ are unforgeable, and not usable by clients other than $C_r$.
Moreover, it must make sure that content authorized for $C_r$ can only be decrypted by 
$C_r$. In the rest of this section we go into the details of how
\KRBCCN\ achieves these requirements and functionalities.

\subsection{Namespace-Based AC Policies}
Instead of traditional service principals in Kerberos, \KRBCCN\ AC policies refer to namespaces, i.e, 
prefixes of content names that correspond to a producer. Recall that a content name is a URI-like 
string composed of arbitrary number of elastic name segments, separated by a \url{`/'} character.
%%GTS: next sentece is poorly worded. Pleae change + clarify!
%%Ivan: I was meaningless. I just removed it.
%In addition, a name segment is always more specific than its predecessor in the content name.
For example, consider a content named:

\centerline{\fbox{\small\bf\noindent\url{/edu/uni-X/ics/cs/students/alice/images/img1.png}}}

The leftmost part, \url{/edu/uni-X/ics/cs}, defines this content's original producer's location.
Subsequent name segments get increasingly specific, defining, e.g., location of the content in a 
directory structure on the producer.

\KRBCCN\ leverages this hierarchical name structure to implement AC policies based on 
content prefixes. For example, to grant Alice permission to retrieve contents under the prefix 
\url{/edu/uni-X/ics/cs/students/alice}, namespace \url{/edu/uni-X/ics/cs/students/alice/*} 
must be included under Alice`s ID in the AC Policy Store, as shown in Figure~\ref{arch}.
This entry would allow Alice to retrieve \url{/edu/uni-X/ics/cs/students/alice/images/img1.png}, 
as well as any content with that same prefix.

Suppose that Bob is a faculty member of the faculty in the same institution and has privileges to 
retrieve contents under own (Bob's) private directory and any content of students' directories.
Bob's entry in the AC Policy store would include two namespaces: 

\centerline{\fbox{\footnotesize\bf\noindent\url{/edu/uni-X/ics/cs/faculty/bob/*} {\rm and} \url{/edu/uni-X/ics/cs/students/*}}}

The former allows Bob to access its own private directory under the 
faculty directory, but no other faculty's private directories.
The latter allows Bob to access contents of any student directory under \url{/edu/uni-X/ics/cs/students/*}.
Finally, suppose that Carl is a system administrator. As such, he has access to all content.
Carl's entry in the AC Policy Store would be the namespace \url{/edu/uni-X/ics/*}, allowing access 
to any content with a name starting with this prefix; this includes all faculty and students' content.

If Alice (who is not yet "logged in", i.e., has no current TGT) wants to issue an interest for 
\url{/edu/uni-X/ics/cs/students/alice/images/img1.png} she first authenticates to \KAS\ to get a TGT.
Alice then uses the TGT to request a CGT from \KTGS\ for namespace \url{/edu/uni-X/ics/cs/students/alice/*}.
Notice that Alice does not need to specify the actual content name -- only the namespace.
Therefore, \KTGS\ does not learn which content Alice wants to retrieve, only the producer's name.
Since CGT is associated with \url{/edu/uni-X/ics/cs/students/alice/*}, it can be used for future interests 
within the same namespace, e.g., \url{/edu/uni-X/ics/cs/students/alice/docs/paper.pdf}.

\subsection{Protocol}\label{protocol}
\begin{table}
\centering
\caption{Notation summary}
\label{notation}
\scalebox{0.7}{
\begin{tabular}{|l|p{9cm}|}
\hline
Notation    			&  Description  							\\ \hline \hline
%%GTS: not used!!!
% $I.name$			&  Name of the issued interest I					\\&\\
$N$				&  A namespace prefix (e.g., edu/uni-X/ics/alice/\*) 				\\&\\
$C_r$			&  Consumer \\ & \\
$TGT\_Name$			&  Ticket-granting ticket name (e.g., edu/uni-X/ics/TGT) that will be routed towards 
					     \KAS \\&\\
$CGT\_Name$			&  Content-granting ticket name (e.g., edu/uni-X/ics/CGT) that will be routed towards \KTGS   \\&\\
$sk_C$      			&  Consumer Secret Key						        \\&\\
$pk_C$			        &  Consumer Public Key, including public UID and certificate        	\\&\\
$k_A$ 	   		 	&  Long-term symmetric key shared between \KAS\ and \KTGS  \\&\\
$k_P$ 	   		 	&  Long-term symmetric key shared between \KTGS\ and a given Content Producer    \\&\\
$s \sample \{0,1\}^{\lambda}$	&  Random ${\lambda}$-bits number generation    	     		\\&\\
$ct \gets Enc_{k}(pt) $		&  Authenticated Encryption of $pt$ using symmetric key $k$		\\&\\
$pt \gets Dec_{k}(ct) $		&  Decryption of $ct$ using symmetric key $k$    	     		\\&\\
$ct \gets Enc_{pk}(pt) $		&  Authenticated encryption of $pt$ using public key $pk$ 		\\&\\
$pt \gets Dec_{sk}(ct) $		&  Decryption of $ct$ using secret key $sk$    	     			\\&\\
\hline
\end{tabular}
}
\end{table}

To retrieve protected content, $C_r$ must go through all of \KRBCCN's three phases, in sequence: 
authentication, authorization, and content retrieval. As discuss below, transition between 
phases is automated on the consumer side, i.e., it requires no extra actions.
Table~\ref{notation} summarizes our notation.\\

\begin{figure*}[!ht]
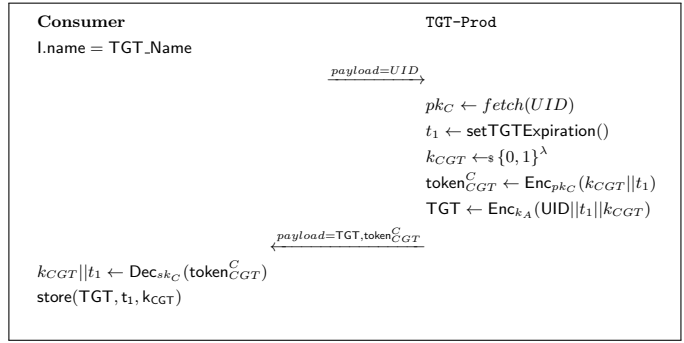

\begin{center}
\fbox{
\scalebox{0.7}{
\procedure{}{%
\textbf{Consumer} \> \> \textbf{\KAS} \\
\mathsf{I.name = TGT\_Name} \> \> \\
\> \xrightarrow{payload = UID} \> \\
\> \> pk_C \gets fetch(UID) \\
\> \> t_1 \gets \mathsf{setTGTExpiration}() \\
\> \> k_{CGT} \sample \{0,1\}^{\lambda} \\
\> \> \mathsf{token}_{CGT}^{C} \gets \mathsf{Enc}_{pk_C}(k_{CGT}||t_1) \\
\> \> \mathsf{TGT} \gets \mathsf{Enc}_{k_{A}}(\mathsf{UID} || t_1 || k_{CGT}) \\
\> \xleftarrow{payload = \mathsf{TGT}, \mathsf{token}_{CGT}^{C}} \> \\
k_{CGT}||t_1 \gets \mathsf{Dec}_{sk_C}(\mathsf{token}_{CGT}^{C}) \> \> \\
\mathsf{store(TGT,t_1,k_{CGT})} \> \> \\
}
}
}
\caption{Consumer authentication protocol}
\label{TGT}
\end{center}
\end{figure*}

\noindent \textbf{Authentication:\\}           
The first phase on \KRBCCN\ verifies consumer identity via authentication.
The authentication protocol in Figure~\ref{TGT} is executed between $C_r$ and \KAS.
If it succeeds, $C_r$ receives a TGT, used in the authorization phase, as proof that $C_r$'s 
identity has been recently verified.

$C_r$ starts by issuing an interest with TGT suffix in the content name (e.g., /uni-X/ics/TGT).
This interest carries as payload consumer's UID, i.e, $C_r$'s username.
Hence, the actual interest name also contains a hash of the payload as its suffix~\footnote{In CCN design, 
an interest carrying a payload must have the hash of the payload appended to its name.}.
The interest is routed by CCN towards \KAS. Upon the interest, \KAS\ looks up $UID$ in its user database 
and retrieves the corresponding public key. The protocol assumes that, when a user enrolls in the system,
a public/private key-pair is generated. Alternatively, a password can be used for the same purpose, as discussed later.
Once \KAS\ successfully locates the user and retrieves the public-key, it proceeds with TGT generation.
Otherwise, it replies with a special error content message indicating unknown user.

TGT is an encrypted structure with three fields: $UID$, $k_{CGT}$, and expiration date $t_1$.
It is encrypted using $k_A$ -- a long-term symmetric key shared between \KAS\ and \KTGS.
Only \KTGS\ can decrypt and access cleartext fields of a TGT. Since $C_r$ needs to present the 
TGT to \KTGS\ during the authorization phase, $UID$ binds the TGT to $C_r$.
This same $UID$ is used later for namespace access rights verification. \KTGS\ uses $t_1$ to verify whether a 
TGT is still valid. TGT expiration time is a realm-specific (and usually realm-wide) parameter reflecting 
the duration of a typical user authenticated session, e.g., 8 hours.
After TGT expires, $C_r$ needs to repeat the authentication protocol with \KAS.
A TGT also contains a short-term symmetric key $k_{CGT}$, encrypted separately for \KTGS\ and $C_r$.
The purpose of $k_{CGT}$ is to allow $C_r$ and \KTGS\ to communicate securely in the subsequent 
authorization protocol phase.
In addition to the TGT, \KAS\ generates $token_{CGT}^{C}$,
which contains the same $t_1$ and $k_{CGT}$ encrypted with the $pk_C$ associated with $UID$.

To transmit the TGT to $C_r$, \KAS\ responds with a content message containing the TGT and 
$token_{CGT}^{C}$, which is routed by CCN back to $C_r$. $C_r$ cannot decrypt, access, or 
modify the TGT due to the use of authenticated encryption.
$C_r$ decrypts $token_{CGT}^{C}$ and caches the TGT for the duration of $t_1$, along with 
$k_{CGT}$. The TGT is presented to \KTGS\ every time $C_r$ needs to request authorization for a
new namespace.\\

\begin{figure*}[!ht]
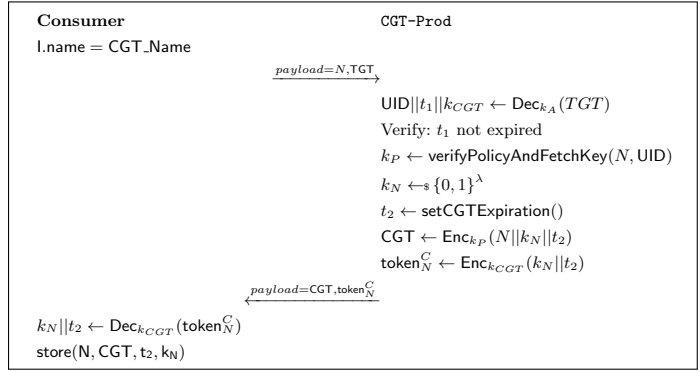

\begin{center}
\fbox{
\scalebox{0.7}{
\procedure{}{%
\textbf{Consumer} \> \> \textbf{\KTGS} \\
\mathsf{I.name = CGT\_Name} \> \> \\
\> \xrightarrow{payload=N, \mathsf{TGT}} \> \\
\> \> \mathsf{UID} || t_1 || k_{CGT} \gets \mathsf{Dec}_{k_{A}}(TGT) \\
\> \> \text{Verify: $t_1$ not expired} \\
\> \> {k_P} \gets \mathsf{verifyPolicyAndFetchKey}(N, \mathsf{UID}) \\ 
\> \> k_{N} \sample \{0,1\}^{\lambda} \\
\> \> t_2 \gets \mathsf{setCGTExpiration}() \\ %% timestamp
\> \> \mathsf{CGT} \gets \mathsf{Enc}_{k_P}( N || k_N || t_2) \\ %% encrypt the key in the ticket
\> \> \mathsf{token}_N^{C} \gets \mathsf{Enc}_{k_{CGT}}(k_N||t_2) \\
\> \xleftarrow{payload= \mathsf{CGT}, \mathsf{token}_N^{C}} \> \\
k_N || t_2 \gets \mathsf{Dec}_{k_{CGT}}(\mathsf{token}_N^{C}) \> \\
\mathsf{store(N,CGT,t_2,k_N)} \>
}
}
}
\caption{Consumer-data authorization protocol}
\label{CGT}
\end{center}
\end{figure*}

\noindent \textbf{Authorization:\\}
The authorization phase (Figure~\ref{CGT}) is executed between $C_r$ and \KTGS.
It requires $C_r$ to have a valid TGT, acquired from the authentication phase described above.
Upon successful completion of the authorization protocol, $C_r$ obtains a namespace-specific 
CGT, which demonstrates $C_r$'s  authorization to access a particular restricted content namespace.
However, the CGT does not reveal $C_r$'s identity to the content producer; $C_r$'s authorization is 
ascertained based on possession of a correct session key.

$C_r$ starts the protocol by sending an interest with name set to \emph{CGT\_Name}.
The payload includes the namespace prefix $N$ (e.g., \url{/edu/uni-X/ics/cs/students/alice/*}) 
for which authorization is being requested and a non-expired TGT.
Optionally, if confidentiality for namespace $N$ is an issue, $C_r$ can compute 
$Enc_{k_{CGT}}(N)$. instead of sending $N$ in clear.
When \KTGS\ receives this interest, it uses $k_A$ (long-term symmetric key shared by \KAS\ and \KTGS) 
to decrypt the TGT and obtains $UID$, $t_1$ and $k_{CGT}$. Next, \KTGS\ checks TGT for expiration. 
It then optionally (if encryption was used in the interest) computes $N \gets Dec_{k_{CGT}}(Enc_{k_{CGT}}(N))$.

If the TGT is successfully verified, \KTGS\ invokes \textit{verifyPolicyAndFetchKey} procedure,
which (1) fetches AC rules for user $UID$; (2) verifies if $N$ is an authorized prefix for $UID$; and 
(3) returns $k_P$ -- symmetric key associated with the producer for $N$. $k_P$ is later used to encrypt the CGT
such that only the appropriate producer can decrypt it.
% If  multiple machines produce content under the namespace $N$, 
% the procedure will return $K=\{k_{P1}, ..., k_{Pn}\}$, i.e., the set of keys associated with each machine.

Similar to a TGT, a CGT carries an expiration $t_2$ and a fresh key $k_N$. The latter is used between 
$C_r$ and the content producer for confidentiality and mutual authentication, as discussed later.
However, instead of $UID$, a CGT includes $N$, i.e., a CGT proves to the content producer that 
whoever possesses $k_N$ is authorized to access content under $N$.
Also, a $token_N^{C} \gets Enc_{k_{CGT}}(k_N||t_2)$ is sent to $C_r$, such that $C_r$ 
can obtain $k_N$ and $t_2$.

In response to a CGT interest, $C_r$ receives a content packet containing the CGT and $token_N^{C}$.
$C_r$ decrypts $token_N^{C}$ using $K_{CGT}$ and creates a cache entry containing: $N$, the CGT, 
$k_N$, and $t_2$. This cached information is used (until time $t_2$) in all future requests for 
content under $N$.\\

\noindent \textbf{Authorized Content Request:\\}
On the consumer (client) side, a \KRBCCN\ content request is similar to a regular CCN interest,
except that $C_r$ needs to include a valid CGT in the payload. An authorized interest name has the format: 
$N||suffix$ (e.g., \url{/edu/uni-X/ics/cs/students/alice/images/img1.png}), where $N$ is authorized by the 
CGT, and $suffix$ specifies which content is being requested under namespace $N$.
Note that, as long as $C_r$ has proper access rights, a single CGT allows accessing 
any content with prefix $N$.

The secure content retrieval phase is in Figure~\ref{req}. When the producer receives an interest for a 
restricted content, it first decrypts the CGT and verifies its expiration.
Note that $k_P$ used to decrypt the CGT is shared between the producer and \KTGS.
Thus, successful decryption (recall that we use authenticated encryption) implies that CGT was 
indeed generated by \KTGS\ and has not been modified.
The producer obtains $k_N$, which is also known to $C_r$.
The producer encrypts requested content using $k_N$, i.e., $D' \gets Enc_{k_N}(D)$. $D'$ is returned to 
$C_r$, which decrypts it to obtain $D$.

Note that, by replaying the interest issued by $C_r$, anyone can retrieve $D'$.
This might not appear problematic since only the authorized consumer (who has $k_N$) can
decrypt $D'$. However, in some application scenarios this might be troublesome, e.g.:
\begin{compactitem}
\item Production of content requires a lot of computation, e.g., expensive encryption.
In this case, an adversary can replay legitimate interests previously issued by authorized consumers.
The adversary's goal might be to mount a DoS attack on the producer.
\item The producer might be a peripheral device, e.g., a printer.  In this setting, the interest might be a request 
to print a (perhaps very large) document and returned content $D$ might be a mere confirmation of it having
been printed. In this case, the replay attack allows the adversary to print the same document multiple times,
resulting in DoS.
\end{compactitem}
This issue occurs since the producer does not authenticate $C_r$ for each interest.
A modified version of the protocol, shown in Figure~\ref{auth_req}, addresses the problem.
It uses a challenge-response protocol that allows the producer to confirm that $C_r$ possesses 
$k_N$ before producing the content or providing service.
As a down-side, this  incurs an additional round of communication for the challenge-response protocol.\\

\begin{figure}[!ht]
\centering
\begin{minipage}{0.45\textwidth}
\fbox{
\scalebox{0.6}{
\procedure{}{%
\textbf{Consumer} \> \> \textbf{Content Producer} \\
\mathsf{I.name = N||suffix} \> \> \\
\> \xrightarrow{payload= CGT} \> \\
\> \>  N' || k_N || t_2 \gets \mathsf{Dec}_{k_P}(CGT) \\
\> \> \text{Verify $N'$ is prefix of I.name} \\
\> \> \text{Verify $t_2$ expiration} \\
\> \> D \gets \mathsf{ProduceData}(\mathsf{I.name}) \\
\> \> D' \gets \mathsf{Enc}_{k_N}(D) \\
\> \xleftarrow{payload=D'} \> \\
D \gets \mathsf{Dec}_{K_N}(\mathsf{D'}) \> \> \\
}
}
}
\caption{Content retrieval \textit{without} optional challenge-response based consumer authentication}
\label{req}
\end{minipage}%
\hspace{0.03\textwidth}
\begin{minipage}{.45\textwidth}
\fbox{
\scalebox{0.6}{
\procedure{}{%
\textbf{Consumer} \> \> \textbf{Content Producer} \\
\mathsf{I.name = N||suffix} \> \> \\
\> \xrightarrow{payload= c_1 , CGT} \> \\
\> \>  N' || k_N || t_2 \gets \mathsf{Dec}_{k_P}(CGT) \\
\> \> \text{Verify $N'$ is prefix of I.name} \\
\> \> \text{Verify $t_2$ expiration} \\
\> \> n_1 \sample \{0,1\}^{\lambda} \\
\> \> chall \gets \mathsf{Enc}_{k_N}(n_1) \\
\> \xleftarrow{payload=chall} \> \\
n_1 \gets \mathsf{Dec}_{K_N}(\mathsf{chall}) \> \> \\
reply \gets \mathsf{Enc}_{k_N}(n_1 + 1)\\
\> \xrightarrow{payload= reply} \> \\
\> \> n_1' \gets \mathsf{Dec}_{K_N}(\mathsf{reply}) \\
\> \> \text{Verify: $(n_1'-1) = n_1$} \\
\> \> D \gets \mathsf{ProduceData}(\mathsf{I.name}) \\
\> \> D' \gets \mathsf{Enc}_{k_N}(D) \\
\> \xleftarrow{payload=D'} \> \\
D \gets \mathsf{Dec}_{K_N}(\mathsf{D'}) \> \> \\
}
}
}
\caption{Content retrieval \textbf{including} optional challenge-response based consumer authentication}
\label{auth_req}
\end{minipage}
\end{figure}

\begin{figure}[!h]
\begin{center}
\includegraphics[width=0.5\textwidth]{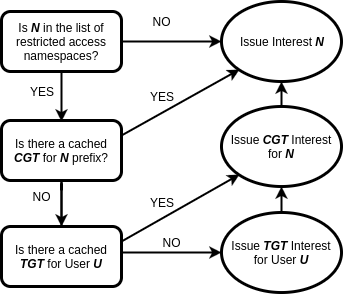}
\caption{\KRBCCN\ work-flow for transparent execution on consumers}
\label{client-workflow}
\end{center}
\end{figure}

\noindent \textbf{Transparent Execution \& Ticket Caching:\\}
Recall that $C_r$ must issue three types of interests, for: authentication, authorization, and 
the actual content request.
This process is transparent to the user since \KRBCCN\ consumer-/client-side code handles these
steps by following the work-flow in Figure~\ref{client-workflow}.

Whenever $C_r$ issues an interest, \KRBCCN\ client intervenes and checks whether
the name is part of any restricted namespace. If so, it looks up the local cache of CGT-s
to find a CGT for prefix $N$. If a valid CGT is found, it is added to the interest payload and 
the interest is issued. A cached and valid CGT can be used to skip the first two phases, 
allowing authenticated and authorized content retrieval in one round.

If no valid cached CGT is found, \KRBCCN\ client looks up a cached TGT.
If a valid TGT is found, the authentication phase is skipped. The client requests a CGT and 
uses it to request the actual content. This process takes two rounds.

In the worst case all three phases are executed, which results in three rounds of communication.
%%GTS: next sentence makes NO SENSE.
%%Ivan: fixed.
Since consumers only request TGT and/or CGT-s when these tickets expire, ticket caching also 
reduces the number of requests (and overall traffic volume) flowing to \KAS\ and \KTGS.
In practice, we expect CGT-s and TGT-s to be long-lived, i.e., on the order of hours or days, 
similar to current single sign-on systems. This means that the bulk of authorized content 
retrieval can be performed in one round.
If mutual authentication (per protocol in Figure~\ref{auth_req}) is demanded by the
producer, one extra round is required.

\section{Implementation \& Performance Evaluation}\label{eval}
This section discusses our \KRBCCN\ prototype implementation and its performance.
\subsection{Methodology}
\KRBCCN\ is implemented as an application service running as specific purpose producers that produce tickets.
Also, consumer-side code is modified to implement the work-flow for authenticated and authorized content
request in Figure~\ref{client-workflow}. Our implementation uses the CCNx software stack~\cite{CCNxGithub} 
and the cryptographic library Sodium~\cite{sodiumGithub}. Both publicly available and written in \textbf{\texttt{C}}. 
For authenticated PKE operations, we use Sodium Sealed-Boxes~\cite{bernstein2006curve25519}, 
implemented over X25519 elliptic curves. AES256-GCM~\cite{dworkin2007recommendation} is used to 
encrypt-then-MAC, i.e., for authenticated symmetric-key encryption.

Experiments presented in this section were ran on an Intel Core i7-3770 octacore CPU @3.40GHz, 
with 16GB of RAM, running Linux (Ubuntu 14.04LTS). Content payload sizes for interests 
were set to 10 kilobytes. Payload sizes of TGT and CGT contents are 228 bytes and 165 bytes, respectively. 
Each carries the respective ticket/token pair, as described in Section~\ref{design}.
In every experiment, each participating entity's process was
assigned as a high priority, and each ran in a single processor core.
Unless stated otherwise, results are an average of 10 executions, presented with 95\% confidence intervals.

Figure~\ref{testbed} presents our network testbed. The goal is to evaluate \KRBCCN's overhead. To avoid 
topology-specific delays, we used a minimal setup containing a single producer $P$, \KAS, and \KTGS.
These entities are interconnected by an unmodified CCNx Athena router. 

%\begin{figure}
%\begin{center}
%\includegraphics[width=0.3\columnwidth]{setup.png}
%\caption{Experimental testbed}
%\label{testbed}
%\end{center}
%\end{figure}

\subsection{Experiments}

\begin{figure}[!ht]
\centering
\begin{minipage}{0.49\textwidth}
  \centering
  \includegraphics[width=0.7\linewidth]{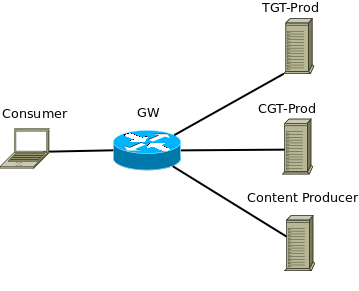}
  \caption{Experimental testbed}
  \label{testbed}
\end{minipage}%
\begin{minipage}{.49\textwidth}
  \centering
  \includegraphics[width=\linewidth]{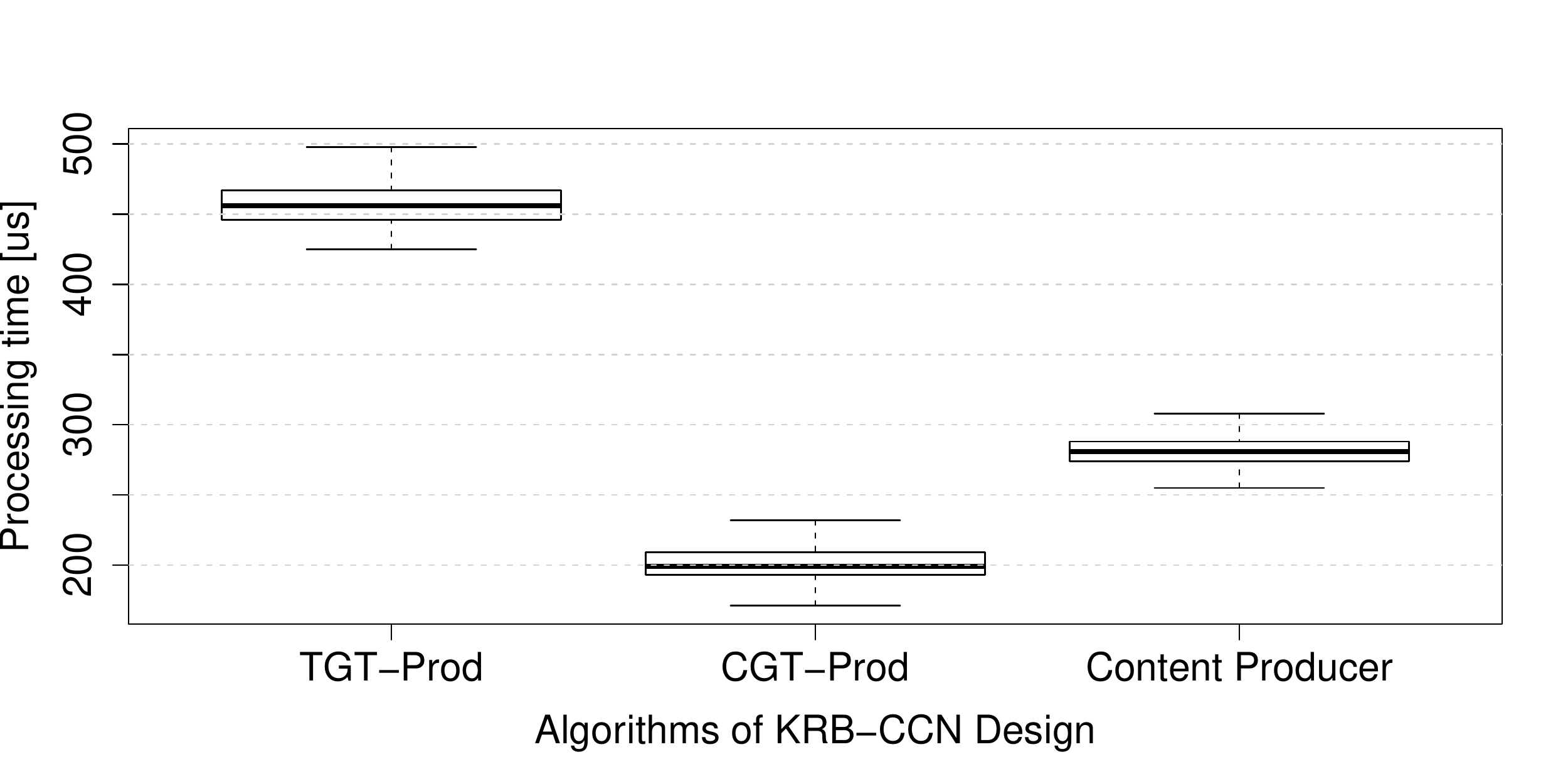}
  \caption{Statistical distribution of the per-interest processing time (in $\mu s$) at each of \KRBCCN\ producers}
  \label{times}
\end{minipage}
\begin{minipage}{.48\textwidth}
  \centering
  \includegraphics[width=\linewidth]{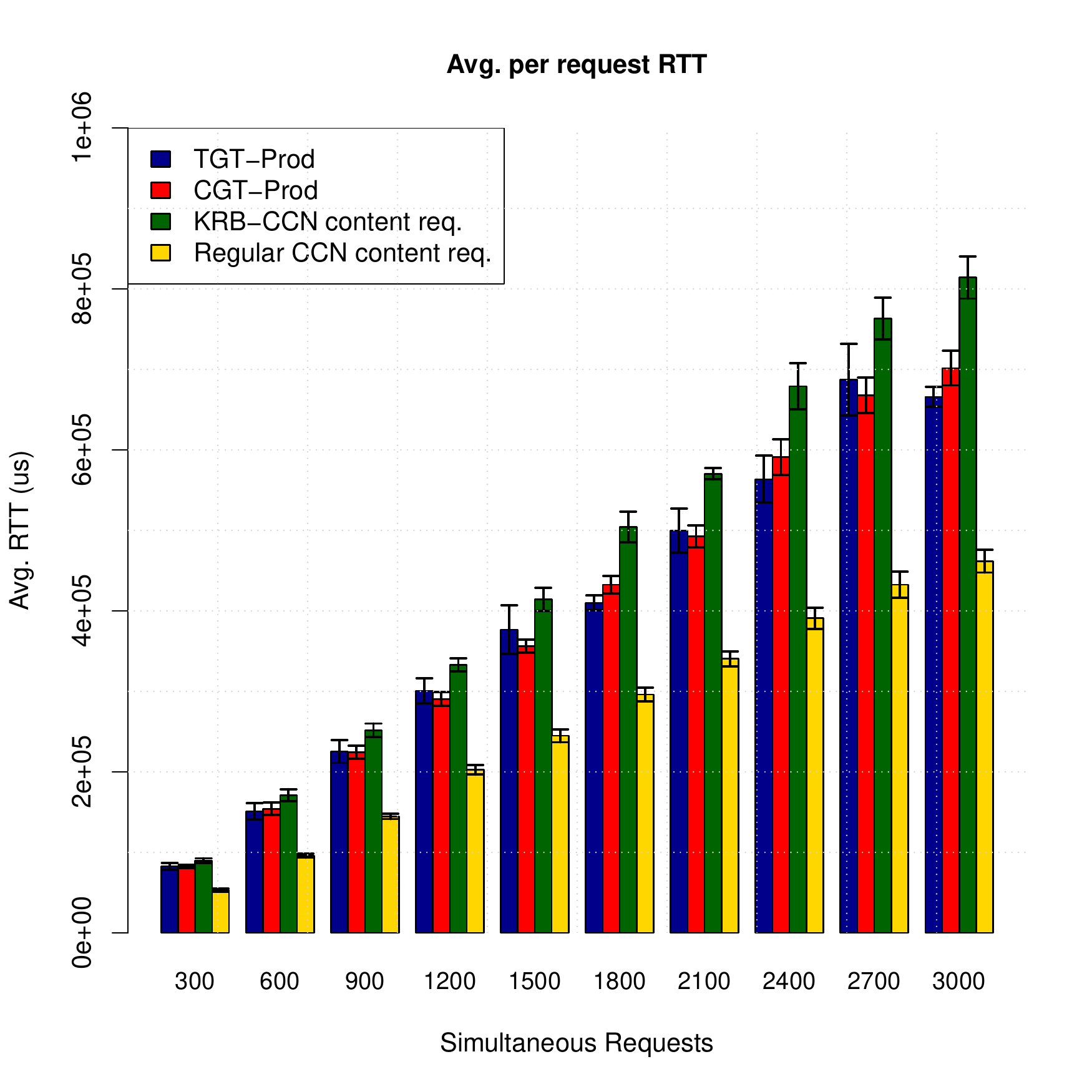}
  \caption{Average RTT per request with massive simultaneous requests to the same producer}
  \label{rtt}
\end{minipage}%
\hspace{0.2cm}
\begin{minipage}{.48\textwidth}
  \centering
  \includegraphics[width=0.95\linewidth]{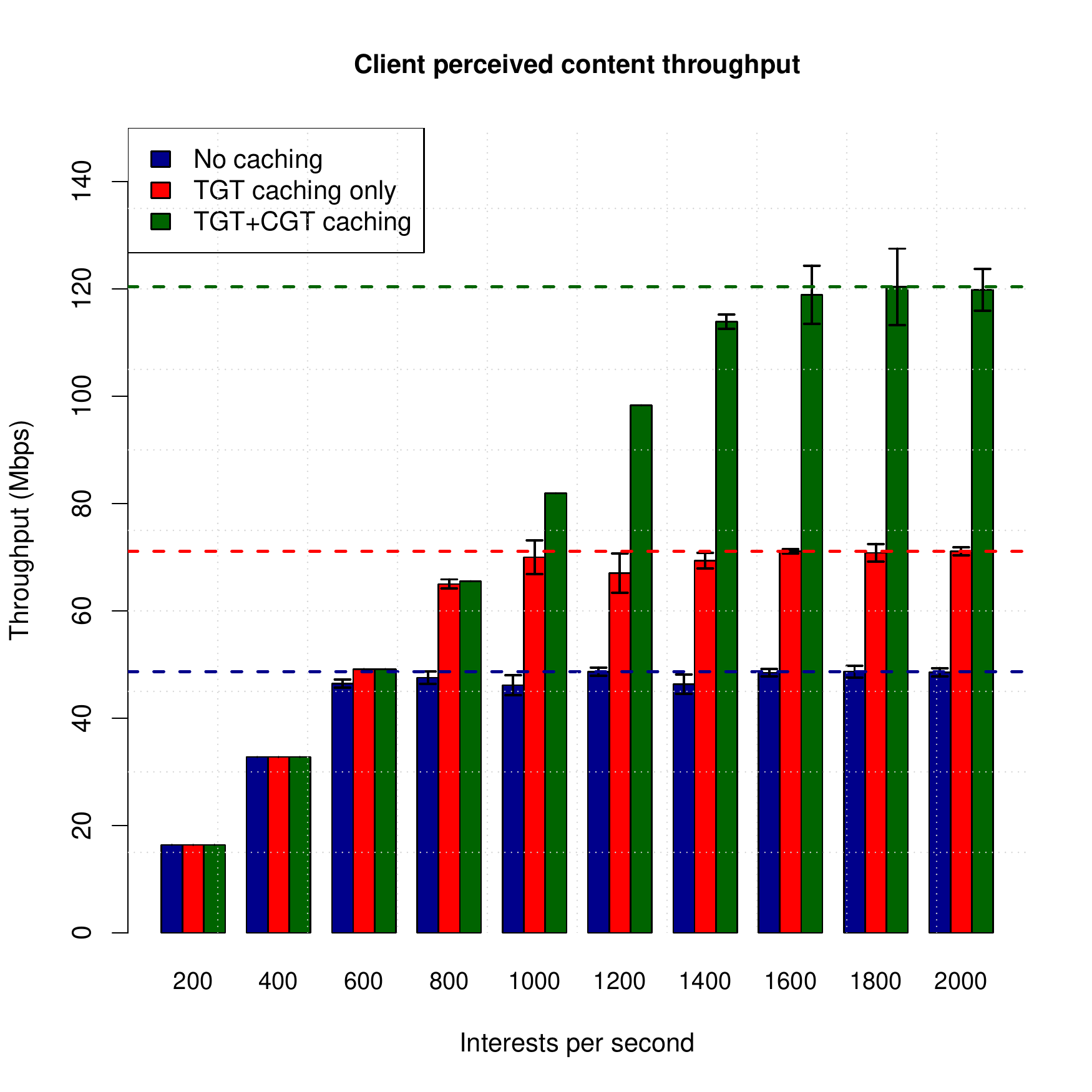}
  \caption{Consumer perceived throughput under different ticket caching policies}
  \label{thput}
\end{minipage}%
\end{figure}

We start by measuring per-request processing times at each producer: \KAS, \KTGS, and $P$. Each of these 
processes was executed $1,000$ times.
Figure~\ref{times} presents the distribution, as box-plots, of processing time for verifying an incoming interest 
and replying with the content (either ticket, or authorized encrypted content) at each producer type.
Figure~\ref{times} shows that the most computationally expensive part is TGT issuance (about $500\mu~s$ per 
request). Higher computational overhead for TGT issuance makes sense because 
the authentication token ($token_{CGT}^{C}$ in Figure~\ref{TGT}) is encrypted with $C_r$'s public key.
%%GTS: Obvious!
% This implies higher overhead due to the use asymmetric-key encryption.
In case of password-based authentication, public key encryption is replaced by much faster 
symmetric key encryption using a password-derived key.
This incurs much lower computational overhead on \KAS.

Times for CGT issuance and content production are around $200\mu~s$ and $300\mu~s$, respectively.
Time is naturally higher for the latter, since encrypted data is larger.
In case of content production, the whole content (10kB) is encrypted.
In a CGT request, only the CGT and the token need to be encrypted, resulting in a faster processing time.

%\begin{figure}
%\begin{center}
%\includegraphics[width=0.3\columnwidth]{times.pdf}
%\caption{Statistical distribution of the per-interest processing time (in $\mu s$) at each of \KRBCCN\ producers}
%\label{times}
%\end{center}
%\end{figure}

To investigate how \KRBCCN\ entities cope under increasing congestion, we flood them with a massive 
number of simultaneous interests: from 300 to 3000. We then measure average Round-Trip Time (RTT) 
per type of issued interest. Figure~\ref{rtt} shows the RTTs for each response type.
We also include the RTT for regular CCN content retrieval.
%%GTS: why do you use "content request"? Isn't it what we call an "interes"???
% The regular content request is an unmodified CCN content request.
Since it incurs no extra processing overhead, the regular content RTT is the lower bound for RTTs in 
CCNx implementation.

%%GTS: horrible prose below... Also, stop using "CCN request"
%%Ivan: tried to fix.
%The RTT of the three types of requests in \KRBCCN\ protocol demonstrate to be affordable even in congested scenarios.
The average RTT for interests for TGT, CGT, and authorized encrypted content are similar.
The latter is slightly higher as more data (10kB per interest) must traverse the reverse path back to the consumer.
\KRBCCN\ requests incur in average $\sim60\%$ higher RTT than unmodified content retrieval.
In the largest scale test case, with 3000 simultaneous interests issued for each producer, content replies are received in less than $800ms$.

%\begin{figure}
%\begin{center}
%\includegraphics[width=0.3\columnwidth]{rtt.pdf}
%\caption{Average RTT per request with massive simultaneous requests to the same producer}
%\label{rtt}
%\end{center}
%\end{figure}

Finally, we also measure the overall throughput perceived by the consumer in three possible scenarios:

\begin{itemize}
\item \textbf{Cached TGT and CGT:} in this case the consumer requests contents under the same namespace.
Therefore, the same (non-expired) cached CGT can be used for authorized content retrieval, allowing the client to skip the authentication and authorization phases.
\item \textbf{Cached TGT only:} this is the case in which no AC ticket caching happens.
It happens when the consumer always requests contents under different namespaces or because the realm owner demands consumers to request a fresh CGT for each content.
In this case only the authentication part of the protocol is skipped at each request.
\item \textbf{No caching:} This is the case in which the realm owner does not allow single sign-on through TGT caching nor authorization reuse through CGT caching.
Three requests (authentication, authorization, and content retrieval) are required for each content packet.
\end{itemize}

Recall that issuance of appropriate interests in each of the cases is automatically handled by \KRBCCN\ client software running on the consumer.

For each of the above cases, we gradually increase the rate of interests requested per second until throughput reaches its asymptotic limit.
By analyzing throughput results, presented in Figure~\ref{thput}, we can observe the benefit of ticket caching.
When both TGT and CGT caching are enabled, the client perceived throughput is higher than in the other cases, as actual content can be retrieved with a single interest.
Conversely, caching only TGT-s is still better than not caching any type of ticket, because in this case the authentication phase can be skipped.

%\begin{figure}
%\begin{center}
%\includegraphics[width=0.3\columnwidth]{thput.pdf}
%\caption{Consumer perceived throughput under different ticket caching policies}
%\label{thput}
%\end{center}
%\end{figure}

\section{Related Work}\label{rw}
%
% \KRBCCN\ is an approach for providing authentication, AC, and confidentiality in CCNs.
Previous related efforts provide other types of security services currently available in IP-based networks.
ANDaNA~\cite{dibenedetto2011andana} is an anonymity service analogous to Tor \cite{tor} 
that uses CCN application-layer onion routing. Mosko, et al.~\cite{mosko2017mobile} proposed a 
TLS-like key exchange protocol for building secure sessions for ephemeral end-to-end security in CCN.
Similar to IPSec-based VPNs~\cite{doraswamy2003ipsec}, CCVPN~\cite{ccvpn} is a network-layer approach for 
providing content confidentiality and interest name privacy via  secure tunnels between physically 
separated private networks.

There are several CCN-based techniques that implement so-called Content-Based Access Control (CBAC).
They tend to rely on content encryption under appropriate consumer keys to enforce AC. A group-based AC scheme 
for early versions of CCN was proposed in~\cite{smetters2010ccnx}. Policies tie groups of consumers to content, 
ensuring that only consumers belonging to authorized groups can decrypt restricted content.
Similarly, Misra et al. \cite{misra2013secure} proposed an AC scheme based on broadcast 
encryption~\cite{bcast1,bcast2}. Wood et al.~\cite{wood2014flexible} proposed several AC schemes based on 
proxy re-encryption~\cite{proxy1,proxy2} to enable consumer personalized content caching.
An attribute-based AC system, using attribute-based cryptography~\cite{abc, bethencourt2007ciphertext} was 
proposed in~\cite{ion2013toward}. CCN-AC~\cite{kuriharay2015encryption} is a framework that unifies CBAC-type 
methods by providing a flexible encryption-based AC framework.  It relies on manifest-based content retrieval 
specification (defined in CCNx 1.0~\cite{solis2014ccn}) to enable flexible and extensible AC policy 
specification and enforcement. A similar approach is proposed in NDN-NBAC~\cite{yu2015name} 
framework. In these frameworks, data owners generate and distribute private keys to consumers via 
out-of-band channels. Producers receive corresponding public keys also via out-of-band channels.
These public keys are used to encrypt one-time (per-content) symmetric keys.

In a different vein, Ghali et al.~\cite{ghali2015interest} proposed an Interest-Based Access Control (IBAC) 
scheme, wherein access to protected content is enforced by making names secret and unpredictable -- based
on encryption with keys known only to authorized consumers.
Compared with CBAC, IBAC has the advantage of preserving interest name privacy and allowing content caching.
However, IBAC must be used in conjunction with CBAC to preclude unauthorized content retrieval via 
replay of previously issued obfuscated interest names.

In all schemes discussed above, authentication, authorization/AC, and confidentiality are often convoluted.
In particular, producers are assumed to be implicitly responsible for authentication and authorization.
This implies dealing with identity management and thus violating consumer privacy.
Moreover, authentication and AC are enforced on a per-content basis which is unscalable and expensive.
To the best of our knowledge \KRBCCN\ is the first comprehensive approach to address these issues by 
(1) separating authentication, authorization and content production among distinct  entities;
and (2) issuing re-usable authentication and authorization tickets for restricted namespaces.

\section{Conclusions}\label{sec:conclusion}
We presented \KRBCCN\ -- a comprehensive design for handling authentication, authorization, and access control in private CCN networks, while preserving consumer privacy.
\KRBCCN\ is transparent to consumers and incurs fairly low overhead.
We analyzed \KRBCCN\ security and assessed its performance based on a prototype implementation.
Experimental results show that \KRBCCN\ is a practical and efficient means of providing multiple 
security services in private (stub AS) CCNs. 

\section*{Acknowledgments}

The authors would like to thank Christopher Wood for fruitful discussions and feedback.
This work was supported by CISCO University Research Award (2017).

\bibliographystyle{IEEEtran}
\bibliography{main}

\newpage

\section*{Appendix A: \KRBCCN\ Security Analysis}\label{security}
We now discuss security of \KRBCCN\ design presented in Section~\ref{design}. We start by introducing the assumed adversary model and then 
split security analysis into 3 parts: user authentication, authorization, and content.

\subsection{Adversary Model}
We consider the worst-case scenario: $C_r$, that does not have a valid CGT or TGT,
wants to fetch certain content. Thus, $C_r$ must engage in \KRBCCN\ authentication (Figure~\ref{TGT}), 
authorization (Figure~\ref{CGT}), and authorized content retrieval (Figure~\ref{req} or~\ref{auth_req}, 
depending on the producers' requirements). \KAS\ and \KTGS\ are trusted parties.\\

\textbf{Adversary Goals and Capabilities:} 
The adversary succeeds if it:
\textbf{(1)} retrieves and decrypts unauthorized content; 
\textbf{(2)} retrieves any session key ($k_{CGT}$, $k_N$) or long-term key ($k_A$, $k_P$); or
\textbf{(3)} forges tickets (TGT or CGT), tokens, or contents, leading $C_r$ to believe that such 
forgeries were generated by honest parties;
We consider a generic adversary $Adv$ that is ubiquitous within a given \KRBCCN\ realm. 
$Adv$ can perform the following actions:
\begin{itemize}
\item \textbf{Compromise and deploy compromised routers:} 
We allow $Adv$ to compromise any number of existing CCN routers on the path between communication 
end-points. Hence, $Adv$ learns all information in the routers and can change it at will. This includes 
changing FIB, PITs and content caches. We also allow $Adv$ to deploy its own compromised routers.
\item \textbf{Eavesdrop, analyze, and replay messages:} $Adv$ can observe all CCN messages. 
Moreover, $Adv$ can record and replay any traffic at will.
\item \textbf{Issue interests and publish content:} $Adv$ can deploy its own malicious consumers and producers. 
Thus, $Adv$ can issue arbitrary interests and produce arbitrary content under its owned namespace.
\end{itemize}
\textit{\textbf{Note:}} Protecting software and hardware of consumers or producers is out of scope for \KRBCCN. 
In particular, hardware and software exploits or malware are not considered in our analysis. Similar to IP-based 
Kerberos, \KRBCCN\ is an authentication and authorization system for secure computing platforms 
communicating over an untrusted (CCN) network.

We analyze \KRBCCN\ security in a top-down fashion. Security of content depends on security of the 
authorization phase, which, in turn, relies on secure user authentication. Therefore, we start by arguing 
that content retrieval is secure as long as authorization is secure. Then, we show that authorization is 
secure if user authentication is secure. Finally, we argue security of authentication by showing that 
only the owner of a certain user identity can gain access to content by claiming such identity.
These notions are formalized bellow.

\subsection{Content Retrieval Security}
Content retrieval security means guaranteeing Property~\ref{prop1} (below) in case of content retrieval 
without consumer authentication. When consumer authentication is required, secure content retrieval 
implies both Properties~\ref{prop1} and~\ref{prop2}.
\begin{prop}\label{prop1}
\textit{
Given an interest containing a CGT in the format \\ $CGT \gets Enc_{k_P}( N || k_N || t_2)$ as its payload, 
assuming that $k_N$ is only known by $C_r$ and $k_P$ is only known by producer $Pr$,  it must hold that:
\begin{enumerate}
%%GTS: what is "integral"???
 \item $P_r$ only replies to valid (non-expired and authentic) CGT-s issued by \KTGS\ 
 and only $P_r$ can decrypt such CGT-s;
 \item A valid CGT for $N$ can not be used to retrieve contents that do not belong to namespace $N$;
 \item Only $C_r$ can decrypt content generated by $P_r$ in response to an interest containing such CGT;
 \item $C_r$ can check content authenticity (i.e., whether content originated at $P_r$) and integrity.
\end{enumerate}
}
\end{prop}
\begin{cl}
\KRBCCN\ content retrieval protocol (Figure~\ref{req}) retains Property~\ref{prop1}. 
\end{cl}

\begin{ps}\label{ps1}
Item (1) is assured because only \KTGS\ knows $k_P$ and authenticated encryption with $k_P$ is used to encrypt CGT.
Moreover, the expiration $t_2$ is checked before issuing content replies.
Item (2) follows from the integrity of received CGT (guaranteed by item 1) and the fact that $P_r$ checks if the received request is for content that has the prefix $N$.
Item (3) is true because content replies are encrypted under $k_N$, which is only known by $C_r$.
Item (4) is assured by the use of authenticated encryption using $k_N$. Only $P_r$ has $k_N$, because $k_N$ is transmitted inside CGT, which is encrypted with $k_P$.
Therefore, $C_r$ can be sure that such content was in fact generated by $P_r$ and also verify the received content integrity.
\end{ps}

\begin{prop}\label{prop2}
\textit{
Given an interest containing a CGT in the format \\ $CGT \gets Enc_{k_P}( N || k_N || t_2)$ as its payload, if $k_N$ is only known by consumer $C_r$ and $k_P$ is only known by producer $Pr$, only $C_r$ is able to use such CGT to get $P_r$
to execute $D \gets \mathsf{ProduceData}(\mathsf{I.name})$.
}
\end{prop}

\begin{cl}
\KRBCCN\ content retrieval protocol with consumer authentication (Figure~\ref{auth_req}) retains Properties~\ref{prop1} and~\ref{prop2}. 
\end{cl}

\begin{ps}
The proof that content retrieval protocol with consumer authentication has Property~\ref{prop1} is equivalent to Proof~\ref{ps1}.
Property~\ref{prop2} is achieved because, in the protocol in Figure~\ref{auth_req}, $P_r$ also plays a challenge response protocol with $C_r$, guaranteeing that
$C_r$ knows $K_N$ before executing $D \gets \mathsf{ProduceData}(\mathsf{I.name})$.
\end{ps}

Both properties assume that $K_P$ is only known to \KTGS\ and $P_r$.
This key is shared between them when $P_r$ enrolls into a \KRBCCN\ realm.
The properties also assume that $K_N$ is only known to consumer $C_r$, which follows from authorization security, discussed in Section~\ref{autho_sec}.

\subsection{Authorization Security}\label{autho_sec}

Authorization security is represented by Property~\ref{prop3}.

\begin{prop}\label{prop3}
\textit{
Given an authorization request interest for namespace $N$ with $TGT \gets \mathsf{Enc}_{k_{A}}(\mathsf{UID} || t_1 || k_{CGT})$ as payload, assuming that $k_{CGT}$ is only known by consumer $C_r$ and $k_A$ is shared between \KTGS\ and \KAS, it must hold that:
\begin{enumerate}
\item \KTGS\ will only reply to valid (non-expired and authentic) TGT-s issued by \KAS\ and only \KTGS\ is able to decrypt such TGT-s;
\item A CGT is only issued if the UID in the TGT has access to namespace $N$;
\item Only $C_r$ is able to retrieve and verify integrity and authenticity of $k_N$;
\end{enumerate}
}
\end{prop}

\begin{cl}
\KRBCCN\ authorization protocol (Figure~\ref{CGT}) retains Property~\ref{prop3}. 
\end{cl}

\begin{ps}
Item (1) follows from the use of authenticated encryption with $k_A$ to encrypt the TGT and from the expiration verification of $t_1$.
Recall that $k_A$ is only shared between \KTGS\ and \KAS.
Item (2) is guaranteed by the integrity of received TGT (item (1)) and by the verification in the AC Policy database performed by \KTGS.
Item (3) holds because $k_N$ is encrypted with $k_{CGT}$, generating $token_N^C$ (and $k_{CGT}$ is only known by $C_r$).
Therefore, only $C_r$ can decrypt $token_N^C$ and obtain $k_N$.
Integrity of $k_N$ is verifiable by $C_r$ due to the use of authenticated encryption to generate $token_N^C$.
\end{ps}

Property~\ref{prop3} ensures that CGT-s are only issued to authorized consumers because only those are able to use $k_{CGT}$ to decrypt $token_N^C$ and retrieve $k_N$.
The only missing link in \KRBCCN\ security is to ensure that if a given $C_r$ has $k_{CGT}$ such $C_r$ is in fact the owner of UID identity  in the TGT.
This is discussed next, in Section~\ref{auth_sec}.

\subsection{Authentication Security}\label{auth_sec}

Authentication security relies on making sure that only the owner of a claimed UID is able to retrieve $k_{CGT}$, i.e., the key included in the TGT issued for UID.
This is stated by the following property:

\begin{prop}\label{prop4}
Given an identity represented by $(UID,sk_C,pk_C)$, where $pk_C$ is a public-key know to \KAS\ and $sk_C$ is the associated secret-key only know to $C_r$ -- the owner of UID.
It must hold that , for an issued $\mathsf{TGT} \gets \mathsf{Enc}_{k_{A}}(\mathsf{UID} || t_1 || k_{CGT})$, only $C_r$ is able to retrieve the key $k_{CGT}$.
Moreover $C_r$ can verify integrity and authenticity of $k_{CGT}$.
\end{prop}

\begin{cl}
\KRBCCN\ authentication protocol (Figure~\ref{TGT}) retains Property~\ref{prop4}. 
\end{cl}

\begin{ps}
Since $\mathsf{token}_{CGT}^{C}$ is encrypted using $pk_C$, only the owner of UID can decrypt it and recover $k_{CGT}$.
Integrity and authenticity of $k_{CGT}$ are verifiable due to the use of authenticated encryption. 
\end{ps}

In the password-based version of \KRBCCN, instead of encrypting under a public key, \KAS\ would generate $token_{CGT}^{C}$ by encrypting with a symmetric key derived from password.
Presumably, only the user that owns $UID$ would know that secret password and be able to generate the same key to decrypt $token_{CGT}^{C}$.
This approach shares the same characteristics and challenges (e.g., password strength, dictionary attacks, password memorability) of any password-based authentication.
Password-based authentication is an option in \KRBCCN\ design. However, discussing specific challenges of password based authentication, though interesting, is not in the scope of the present work.

\subsection{Discussion}
By retaining Properties~\ref{prop1}, \ref{prop2}, \ref{prop3}, and~\ref{prop4}, \KRBCCN\ ensures secure authentication, AC, content integrity, and content confidentiality.
As \KRBCCN\ protocol runs on consumers and producers, compromised routers and/or eavesdroppers are not able to violate what is guaranteed by the aforementioned properties.   

Replay attacks and spoofed messages do not allow $Adv$ to retrieve (unencrypted) content.
If the content production is heavyweight, DoS via replay can be ruled out by enforcing mandatory mutual authentication (protocol in Figure~\ref{auth_req}).

In addition to the security services discussed in this section, \KRBCCN\ preserves consumer privacy.
Producers only need to verify CGT-s authenticity and integrity, instead of consumers' identities.
CGT-s do not carry UIDs, but instead associated keys, allowing the system to remain secure while preserving privacy.
CGT-s are issued for namespaces (and not for complete content names).
Thus, not even \KTGS\ (which knows the consumer UID) is able to predict which content will be requested by a consumer after issuing a CGT.

With ticket (TGT and CGT) caching, authentication and authorization can be skipped for subsequent interest within the same namespace.
Therefore, most of the times \KRBCCN\ only requires low-cost (symmetric key) cryptographic operations from consumers/producers.

\end{document}